 \documentclass[10pt]{article}
 
 \usepackage{amsmath}
 \usepackage{amssymb}
 
 \usepackage{graphicx}
 
 \usepackage{cite}
 
 \usepackage{color} 
 
 \usepackage[labelfont=bf,labelsep=period,justification=raggedright]{caption}
 
 \makeatletter
 \renewcommand{\@biblabel}[1]{\quad#1.}
 \makeatother

 \date{}
 
\begin{document}
 
 \begin{flushleft}
 {\Large
 \textbf{powerlaw: a Python package for analysis of heavy-tailed distributions}
 }
 \\
 Jeff Alstott$^{1,2\ast}$, Ed Bullmore$^{2}$, Dietmar Plenz$^{1}$
 \\
\bf{1} Section on Critical Brain Dynamics, National Institute of Mental Health, Bethesda, Maryland, USA
 \\
\bf{2} Brain Mapping Unit, Behavioural and Clinical Neuroscience Institute, University of Cambridge, Cambridge, UK
 \\
 $\ast$ E-mail: alstottjd@mail.nih.gov
 \end{flushleft}
 
 \section*{Abstract}
 Power laws are theoretically interesting probability distributions that are also frequently used to describe empirical data. In recent years effective statistical methods for fitting power laws have been developed, but appropriate use of these techniques requires significant programming and statistical insight. In order to greatly decrease the barriers to using good statistical methods for fitting power law distributions, we developed the \verb$powerlaw$ Python package. This software package provides easy commands for basic fitting and statistical analysis of distributions. Notably, it also seeks to support a variety of user needs by being exhaustive in the options available to the user. The source code is publicly available and easily extensible.
 
 \section*{Introduction}
 Power laws are probability distributions with the form:
 \begin{equation}
 p(x) \propto x^{-\alpha}
 \end{equation}
 
 Power law probability distributions are theoretically interesting due to being  "heavy-tailed", meaning the right tails of the distributions still contain a great deal of probability. This heavy-tailedness can be so extreme that the standard deviation of the distribution can be undefined (for $\alpha<3$), or even the mean (for $\alpha<2$). These qualities make for a scale-free system, in which all values are expected to occur, without a characteristic size or scale. Power laws have been identified throughout nature, including in astrophysics, linguistics, and neuroscience \cite{Michel2011, Zipf1935, Beggs2003, Shriki2013}. However, accurately fitting a power law distribution to empirical data, as well as measuring the goodness of that fit, is non-trivial. Furthermore, empirical data from a given domain likely comes with domain-specific considerations that should be incorporated into the statistical analysis.
 
In recent years several statistical methods for evaluating power law fits have been developed \cite{Clauset2009, Klaus2011}. We here introduce and describe \verb$powerlaw$, a Python package for easy implementation of these methods. The \verb$powerlaw$ package is an advance over previously available software because of its ease of use, its exhaustive support for a variety of probability distributions and subtypes, and its extensibility and maintainability. The incorporation of numerous distribution types and fitting options is of central importance, as appropriate fitting of a distribution to data requires consideration of multiple aspects of the data, without which fits will be inaccurate. The easy extensibility of the code base also allows for future expansion of \verb$powerlaw$'s capabilities, particularly in the form of users adding new theoretical probability distributions for analysis.
 
In this report we describe the structure and use of \verb$powerlaw$. Using \verb$powerlaw$, we will give examples of fitting power laws and other distributions to data, and give guidance on what factors and fitting options to consider about the data when going through this process.
 
\begin{figure}[!ht]
\begin{center}
\includegraphics[width=4in]{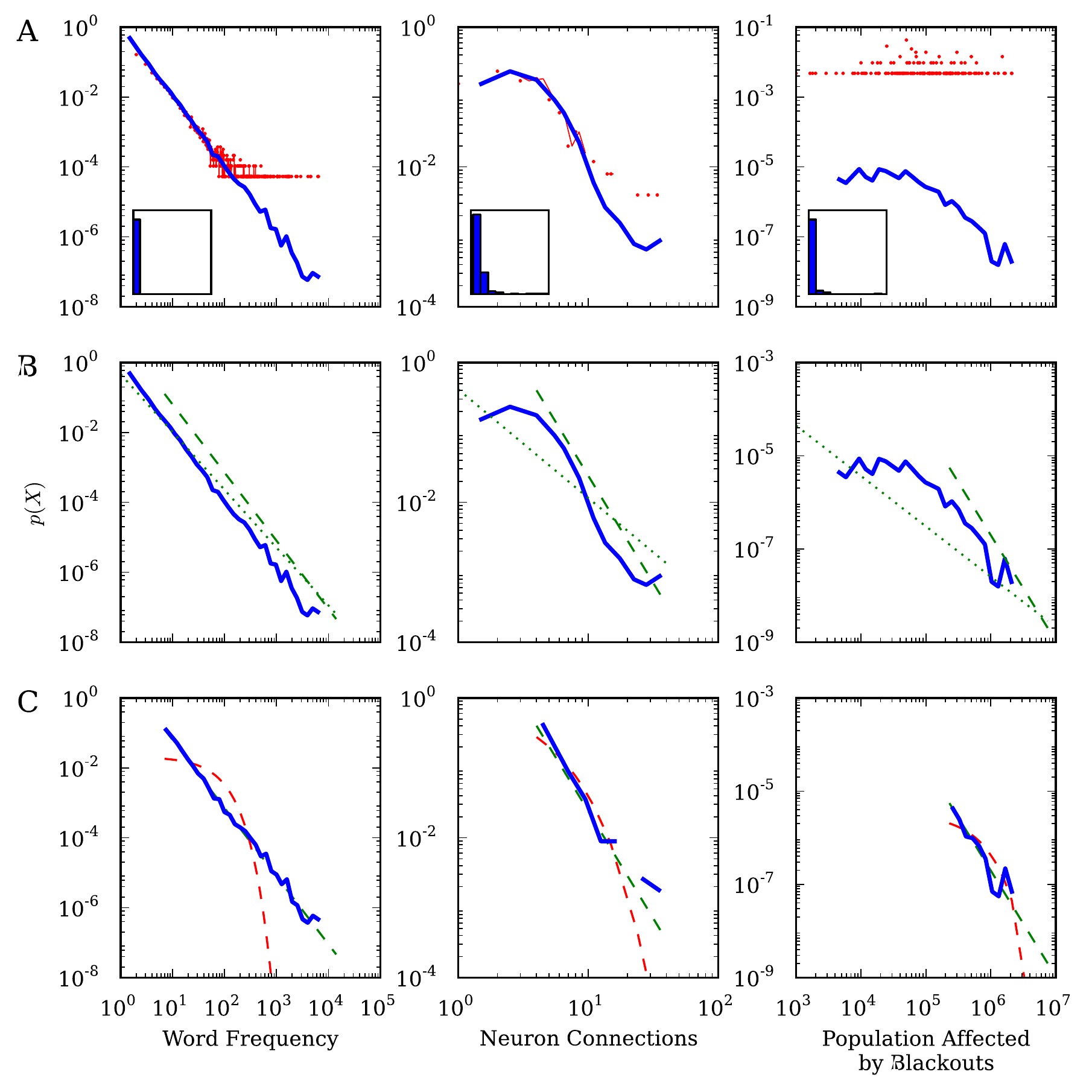}
\end{center}
\caption{
{\bf Basic steps of analysis for heavy-tailed distributions: visualizing, fitting, and comparing.} Example data for power law fitting are a good fit (left column), medium fit (middle column) and poor fit (right column). Data and methods described in text.
a) Visualizing data with probability density functions. A typical histogram on linear axes (insets) is not helpful for visualizing heavy-tailed distributions. On log-log axes, using logarithmically spaced bins is necessary to accurately represent data (blue line). Linearly spaced bins (red line) obscure the tail of the distribution (see text).
b) Fitting to the tail of the distribution. The best fit power law may only cover a portion of the distribution's tail. Dotted green line: power law fit starting at $x_{min}$=1. Dashed green line: power law fit starting from the optimal $x_{min}$ (see Basic Methods: Identifying the Scaling Range). 
c) Comparing the goodness of fit. Once the best fit to a power law is established, comparison to other possible distributions is necessary.  Dashed green line: power law fit starting from the optimal $x_{min}$. Dashed red line: exponential fit starting from the same $x_{min}$.
}
\label{Workflow}
\end{figure}
Figure \ref{Workflow} shows the basic elements of visualizing, fitting, and evaluating heavy-tailed distributions. Each component is described in further detail in subsequent sections. Three example datasets are included in Figure \ref{Workflow} and the \verb$powerlaw$ code examples below, representing a good power law fit, a medium fit, and a poor fit, respectively. The first, best fitting dataset is perhaps the best known and solid of all power law distributions: the frequency of word usage in the English language \cite{Zipf1935}. The specific data used is the frequency of word usage in Herman Melville's novel "Moby Dick" \cite{Newman2005}. The second, moderately fitting dataset is the number of connections each neuron has in the nematode worm \textit{C. elegans} \cite{Towlson2013,Varshney2011}. The last, poorly fitting data is the number of people in the United States affected by electricity blackouts between 1984 and 2002 \cite{Newman2005}. 
 
 Figure \ref{Workflow}A shows probability density functions of the three example datasets. Figure \ref{Workflow}B shows how only a portion of the distribution's tail may follow a power law. Figure \ref{Workflow}C shows how the goodness of the power law fit should be compared to other possible distributions, which may describe the data just as well or better.
 
 The \verb$powerlaw$ package will perform all of these steps automatically. Below is an example of basic usage of \verb$powerlaw$, with explanation following. Using the populations affected by blackouts:
 
 \begin{verbatim}
 > import powerlaw
 > fit = powerlaw.Fit(data)
 Calculating best minimal value for power law fit
 > fit.power_law.alpha
   2.273
 > fit.power_law.sigma
   0.167
 > fit.distribution_compare('power_law', 'exponential')
   (12.755, 0.152)
 \end{verbatim}
 
 An IPython Notebook and raw Python file of all examples is included in Supporting Information.
 
 The design of \verb$powerlaw$ includes object-oriented and functional elements, both of which are available to the user. The object-oriented approach requires the fewest lines of code to use, and is shown here. The \verb$powerlaw$ package is organized around two types of objects, \verb$Fit$ and \verb$Distribution$. The \verb$Fit$ object (\verb$fit$ above) is a wrapper around a dataset that creates a collection of \verb$Distribution$ objects fitted to that dataset. A \verb$Distribution$ object is a maximum likelihood fit to a specific distribution. In the above example, a power law \verb$Distribution$ has been created automatically (\verb$power_law$), with the fitted $\alpha$ parameter \verb$alpha$ and its standard error \verb$sigma$. The \verb$Fit$ object is what the user mostly interacts with.
 
 \section*{Basic Methods}
 \subsection*{Visualization}
 The \verb$powerlaw$ package supports easy plotting of the probability density function (PDF), the cumulative distribution function (CDF; $p(X<x)$) and the complementary cumulative distribution function (CCDF; $p(X\geq x)$, also known as the survival function). The calculations are done with the functions \verb$pdf$, \verb$cdf$, and \verb$ccdf$, while plotting commands are \verb$plot_pdf$, \verb$plot_cdf$, and \verb$plot_ccdf$. Plotting is performed with matplotlib (see Dependencies, below), and \verb$powerlaw$'s commands accept matplotlib keyword arguments. Figure \ref{Workflow}A visualizes PDFs of the example data.
 
 \begin{verbatim}
 > powerlaw.plot_pdf(data, color='b')
 \end{verbatim}
 
 
PDFs require binning of the data, and when presenting a PDF on logarithmic axes the bins should have logarithmic spacing (exponentially increasing widths). Although linear bins maintain a high resolution over the entire value range, the greatly reduced probability of observing large values in the distributions prevents a reliable estimation of their probability of occurrence. This is compensated for by using logarithmic bins, which increases the likelihood of observing a range of values in the tail of the distribution and normalizing appropriately for that increase in bin width. Logarithmic binning is \verb$powerlaw$'s default behavior, but linearly spaced bins can also be dictated with the \verb"linear_bins=True" option. Figure \ref{Workflow}A shows how the choice of logarithmic over linear bins can greatly improve the visualization of the distribution of the data. The blackouts data shows a particularly severe example, in which the sparsity of the data leads individual linear bins to have very few data points, including empty bins. The larger logarithmic bins incorporate these empty regions of the data to create a more useful visualization of the distribution's behavior.
 
 \begin{verbatim}
 > powerlaw.plot_pdf(data, linear_bins=True, color='r')
 \end{verbatim}
 
 As CDFs and CCDFs do not require binning considerations, CCDFs are frequently preferred for visualizing a heavy-tailed distribution. 
 However, if the probability distribution has peaks in the tail this will be more obvious when visualized as a PDF than as a CDF or CCDF. PDFs and CDF/CCDFs also have different behavior if there is an upper bound on the distribution (see Identifying the Scaling Range, below).
 
\begin{figure}[!ht]
\begin{center}
\includegraphics[width=4in]{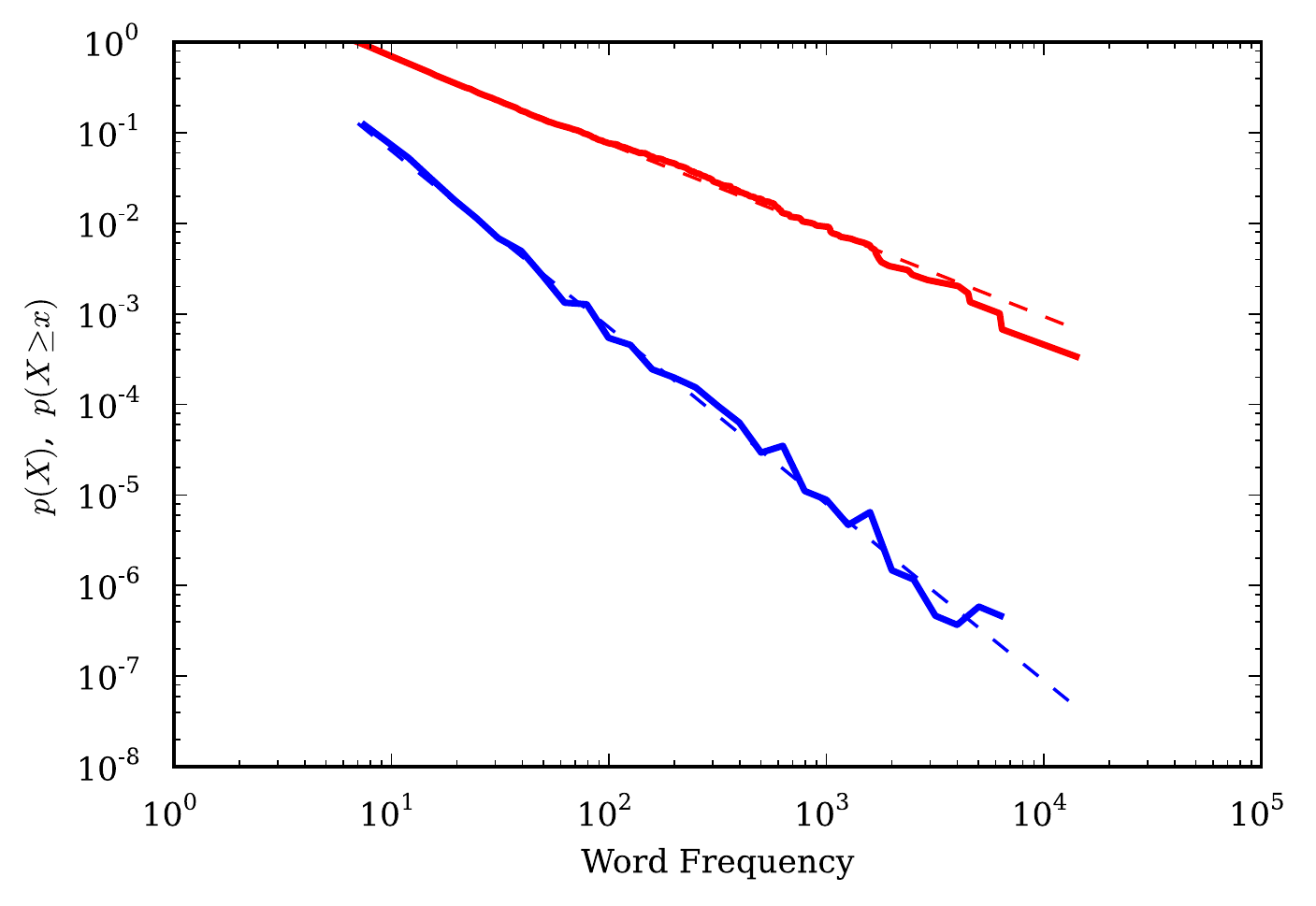}
\end{center}
\caption{
{\bf Probability density function ($p(X)$, blue) and complemenatary cumulative distribution function ($p(X\geq x)$, red) of word frequencies from "Moby Dick".}
}
\label{CCDF}
\end{figure}

 Individual \verb$Fit$ objects also include functions for \verb$pdf$, \verb$plot_pdf$, and their CDF and CCDF versions. The theoretical PDF, CDF, and CCDFs of the constituent \verb$Distribution$ objects inside the \verb$Fit$ can also be plotted. These are useful for visualizing just the portion of the data using for fitting to the distribution (described below). To send multiple plots to the same figure, pass the matplotlib axes object with the keyword \verb$ax$.  Figure \ref{CCDF} shows the CCDF and PDF of the neuron connections dataset and its power law fit. Note that a CCDF scales at $\alpha-1$, hence the shallower appearance.
 
 \begin{verbatim}
 > fig2 = fit.plot_pdf(color='b', linewidth=2)
 > fit.power_law.plot_pdf(color='b', linestyle='--', ax=fig2)
 > fit.plot_ccdf(color='r', linewidth=2, ax=fig2)
 > fit.power_law.plot_ccdf(color='r', linestyle='--', ax=fig2)
 \end{verbatim}
 
 PDF, CDF, and CCDF information are also available outside of plotting. \verb"Fit" objects return the probabilities of the fitted data and either the sorted data (\verb"cdf") or the bin edges (\verb"pdf"). \verb"Distribution" objects return just the probabilities of the data given. If no data is given, all the fitted data is used.
 
 \begin{verbatim}
 > x, y = fit.cdf()
 > bin_edges, probability = fit.pdf()
 > y = fit.lognormal.cdf(data=[300, 350])
 > y = fit.lognormal.pdf()
 \end{verbatim}
 
 \subsection*{Identifying the Scaling Range}
 The first step of fitting a power law is to determine what portion of the data to fit. A heavy-tailed distribution's interesting feature is the tail and its properties, so if the initial, small values of the data do not follow a power law distribution the user may opt to disregard them. The question is from what minimal value $x_{min}$ the scaling relationship of the power law begins. The methods of \cite{Clauset2009} find this optimal value of $x_{min}$ by creating a power law fit starting from each unique value in the dataset, then selecting the one that results in the minimal Kolmogorov-Smirnov distance, $D$, between the data and the fit. If the user does not provide a value for $x_{min}$, \verb$powerlaw$ calculates the optimal value when the \verb$Fit$ object is first created. 
 
 As power laws are undefined for $x=0$, there must be some minimum value. Thus, even if a given dataset brings with it domain-specific reasoning that the data must follow a power law across its whole range, the user must still dictate an $x_{min}$. This could be a theoretical minimum, a noise threshold, or the minimum value observed in the data. Figure \ref{Workflow}B visualizes the difference in fit between assigning $x_{min}=1$ and finding the optimal $x_{min}$ by minimizing $D$. The following \verb$powerlaw$ example uses the blackout data:
 
 \begin{verbatim}
 > fit = powerlaw.Fit(data)
 Calculating best minimal value for power law fit
 > fit.xmin
   230.000
 > fit.fixed_xmin
   False
 > fit.power_law.alpha
   2.273
 > fit.power_law.D
   0.061
 > fit = powerlaw.Fit(data, xmin=1.0)
 > fit.xmin
   1.0
 > fit.fixed_xmin
   True
 > fit.power_law.alpha
   1.220
 > fit.power_law.D
   0.376
 \end{verbatim}
 
 The search for the optimal $x_{min}$ can also be restricted to a range, given as a tuple or list:
 \begin{verbatim}
 > fit = powerlaw.Fit(data, xmin=(250.0, 300.0))
 Calculating best minimal value for power law fit
 > fit.fixed_xmin
   False
 > fit.given_xmin
   (250.000, 300.000)
 > fit.xmin
   272.0
 \end{verbatim}
 
 In some domains there may also be an expectation that the distribution will have a precise upper bound, $x_{max}$. An upper limit could be due a theoretical limit beyond which the data simply cannot go (ex. in astrophysics, a distribution of speeds could have an upper bound at the speed of light). An upper limit could also be due to finite-size scaling, in which the observed data comes from a small subsection of a larger system. The finite size of the observation window would mean that individual data points could be no larger than the window, $x_{max}$, though the greater system would have larger, unobserved data (ex. in neuroscience, recording from a patch of cortex vs the whole brain). Finite-size effects can be tested by experimentally varying the size of the observation window (and $x_{max}$) and determining if the data still follows a power law with the new $x_{max}$ \cite{Beggs2003, Shriki2013}. The presence of an upper bound relies on the nature of the data and the context in which it was collected, and so can only be dictated by the user. Any data above $x_{max}$ is ignored for fitting. 
 
 \begin{verbatim}
 > fit = powerlaw.Fit(data, xmax=10000.0)
 Calculating best minimal value for power law fit
 > fit.xmax
   10000.0
 > fit.fixed_xmax
   True
 \end{verbatim}
 
 For calculating or plotting CDFs, CCDFs, and PDFs, by default \verb"Fit" objects only use data above $x_{min}$ and below $x_{max}$ (if present). The \verb$Fit$ object's plotting commands can plot all the data originally given to it with the keyword \verb$original_data=True$. The constituent \verb$Distribution$ objects are only defined within the range of $x_{min}$ and $x_{max}$, but can plot any subset of that range by passing specific data with the keyword \verb$data$. 
 
\begin{figure}[!ht]
\begin{center}
\includegraphics[width=4in]{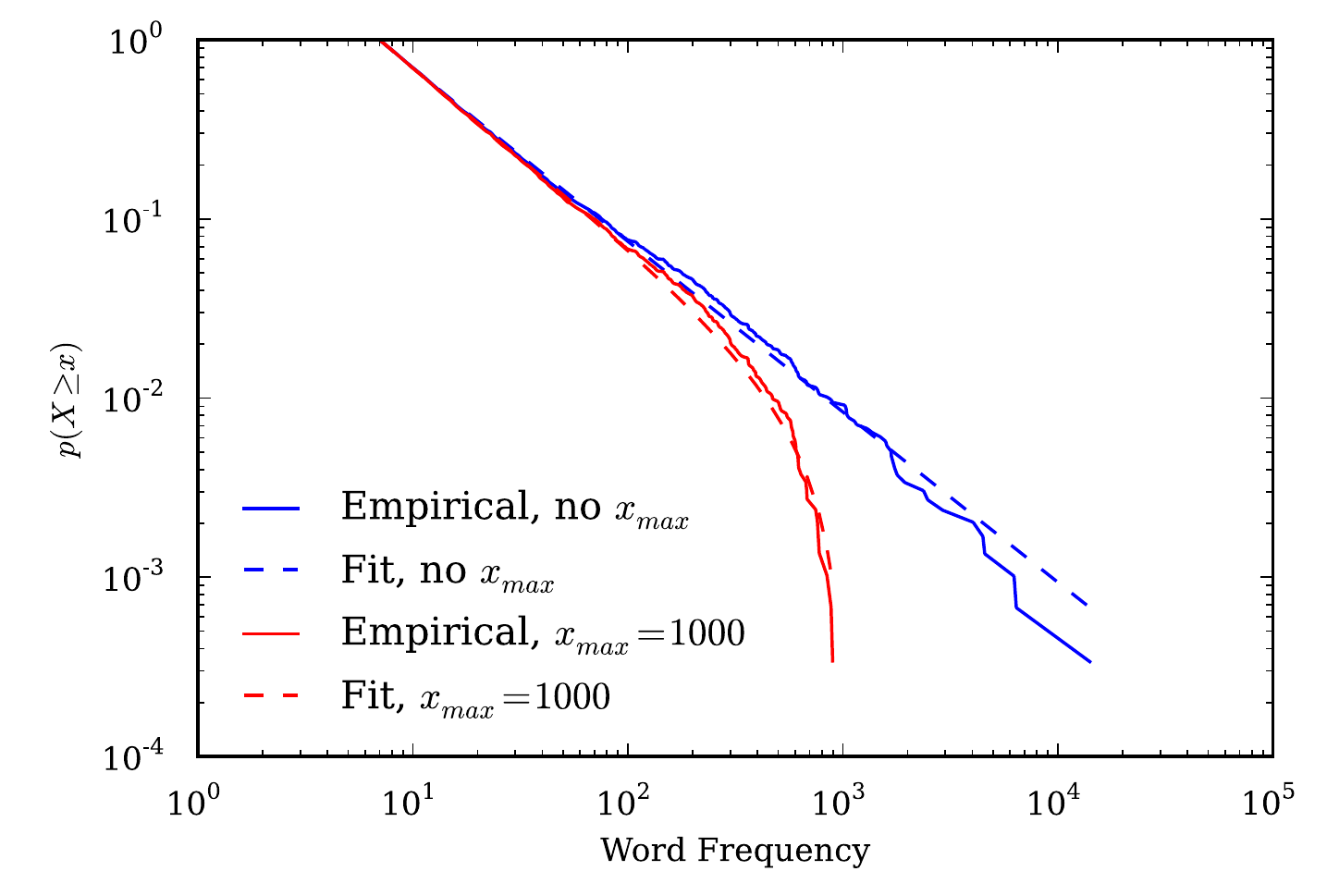}
\end{center}
\caption{
{\bf Complemenatary cumulative distribution functions of the empirical word frequency data and fitted power law distribution, with and without an upper limit $x_{max}$}
}
\label{CCDFmax}
\end{figure}

 When using an $x_{max}$, a power law's CDF and CCDF do not appear in a straight line on a log-log plot, but bend down as the $x_{max}$ is approached (Figure \ref{CCDFmax}). The PDF, in contrast, appears straight all way to $x_{max}$. Because of this difference PDFs are preferrable when visualing data with an $x_{max}$, so as to not obscure the scaling.
 
 \subsection*{Continuous vs. Discrete Data}
 Datasets are treated as continuous by default, and thus fit to continuous forms of power laws and other distributions. Many data are discrete, however. Discrete versions of probability distributions cannot be accurately fitted with continuous versions \cite{Clauset2009}. Discrete (integer) distributions, with proper normalizing, can be dictated at initialization:
 
 \begin{verbatim}
 > fit = powerlaw.Fit(data, xmin=230.0)
 > fit.discrete
   False
 > fit = powerlaw.Fit(data, xmin=230.0, discrete=True)
 > fit.discrete 
   True
 \end{verbatim}
 
 Discrete forms of probability distributions are frequently more difficult to calculate than continuous forms, and so certain computations may be slower. However, there are faster estimations for some of these calculations. Such opportunities to estimate discrete probability distributions for a computational speed up are described in later sections.
 
 \section*{Comparing Candidate Distributions}
 From the created \verb"Fit" object the user can readily access all the statistical analyses necessary for evaluation of a heavy-tailed distribution. Within the \verb"Fit" object are individual \verb"Distribution" objects for different possible distributions. Each \verb"Distribution" has the best fit parameters for that distribution (calculated when called), accessible both by the parameter's name or the more generic "parameter1". Using the blackout data:
 
 \begin{verbatim}
 >  fit.power_law
   <powerlaw.Power_Law at 0x301b7d0>
 > fit.power_law.alpha
   2.273
 > fit.power_law.parameter1
   2.273 
 > fit.power_law.parameter1_name
 > fit.lognormal.mu
   0.154
 > fit.lognormal.parameter1_name
   'mu'
 > fit.lognormal.parameter2_name
   'sigma'
 > fit.lognormal.parameter3_name == None
   True
 \end{verbatim}
 
 The goodness of fit of these distributions must be evaluated before concluding that a power law is a good description of the data. The goodness of fit for each distribution can be considered individually or by comparison to the fit of other distributions (respectively, using bootstrapping and the Kolmogorov-Smirnov test to generate a p-value for an individual fit vs. using loglikelihood ratios to identify which of two fits is better) \cite{Clauset2009}. There are several reasons, both practical and philosophical, to focus on the latter, comparative tests. 
 
 Practically, bootstrapping is more computationally intensive and loglikelihood ratio tests are faster. Philosophically, it is frequently insufficient and unnecessary to answer the question of whether a distribution "really" follows a power law. Instead the question is whether a power law is the best description available. In such a case, the knowledge that a bootstrapping test has passed is insufficient; bootstrapping could indeed find that a power law distribution would produce a given dataset with sufficient likelihood, but a comparative test could identify that a lognormal fit could have produced it with even greater likelihood. On the other hand, the knowledge that a bootstrapping test has failed may be unnecessary; real world systems have noise, and so few empirical phenomena could be expected to follow a power law with the perfection of a theoretical distribution. Given enough data, an empirical dataset with any noise or imperfections will always fail a bootstrapping test for any theoretical distribution. If one keeps absolute adherence to the exact theoretical distribution, one can enter the tricky position of passing a bootstrapping test, but only with few enough data \cite{Klaus2011}. 
 
 Thus, it is generally more sound and useful to compare the fits of many candidate distributions, and identify which one fits the best. Figure \ref{Workflow}C visualizes the differences in fit between power law and exponential distribution. The goodness of these distribution fits can be compared with  \verb$distribution_compare$. Again using the blackout data:
 
 \begin{verbatim}
 > R, p = fit.distribution_compare('power_law', 'exponential', normalized_ratio=True)
 > print R, p
 1.431 0.152
 \end{verbatim}
 
 \verb$R$ is the loglikelihood ratio between the two candidate distributions. This number will be positive if the data is more likely in the first distribution, and negative if the data is more likely in the second distribution. The significance value for that direction is \verb$p$. The \verb$normalized_ratio$ option normalizes \verb$R$ by its standard deviation, $R/(\sigma \sqrt{n})$. The normalized ratio is what is directly used to calculate \verb$p$.
 
 The exponential distribution is the absolute minimum alternative candidate for evaluating the heavy-tailedness of the distribution. The reason is definitional: the typical quantitative definition of a "heavy-tail" is that it is not exponentially bounded \cite{Asmussen2003}. Thus if a power law is not a better fit than an exponential distribution (as in the above example) there is scarce ground for considering the distribution to be heavy-tailed at all, let alone a power law. 
 
 However, the exponential distribution is, again, only the minimum alternative candidate distribution to consider when describing a probability distribution. The fit object contains a list of supported distributions in \verb$fit.supported_distributions$. Any of these distribution names can be used by \verb$distribution_compare$. Users who want to test unsupported distributions can write them into \verb$powerlaw$ in a straightforward manner described in the source code. Among the supported distributions is the exponentially truncated power law, which has the power law's scaling behavior over some range but is truncated by an exponentially bounded tail. There are also many other heavy-tailed distributions that are not power laws, such as the lognormal or the stretched exponential (Weibull) distributions. Given the infinite number of possible candidate distributions, one can again run into a problem similar to that faced by bootstrapping: There will always be another distribution that fits the data better, until one arrives at a distribution that describes only the exact values and frequencies observed in the dataset (overfitting). Indeed, this process of overfitting can begin even with very simple distributions; while the power law has only one parameter to serve as a degree of freedom for fitting, the truncated power law and the alternative heavy-tailed distributions have two parameters, and thus a fitting advantage.  The overfitting scenario can be avoided by incorporating generative mechanisms into the candidate distribution selection process.
 
 \subsection*{Generative Mechanisms}
 The observed data always come from a particular domain, and in that domain generative mechanisms created the observed data. If there is a plausible domain-specific mechanism for creating the data that would yield a particular candidate distribution, then that candidate distribution should be considered for fitting. If there is no such hypothesis for how a candidate distribution could be created there is much less reason to use it to describe the dataset. 
 
 As an example, the number of connections per neuron in the nematode worm \textit{C. elegans} has an apparently heavy-tailed distribution (Figure \ref{Workflow}, middle column). A frequently proposed mechanism for creating power law distributions is preferential attachment, a growth model in which "the rich get richer". In this domain of \textit{C. elegans}, neurons with large number of connections could plausibly gain even more connections as the organism grows, while neurons with few connections would have difficulty getting more. Preferential attachment mechanisms produce power laws, and indeed the power law is a better fit than the exponential:
 
 \begin{verbatim}
 > fit.distribution_compare('power_law', 'exponential')
 (16.384, 0.024)
 \end{verbatim}
 
 However, the worm has a finite size and a limited number of neurons to connect to, so the rich cannot get richer forever. There could be a gradual upper bounding effect on the scaling of the power law. An exponentially truncated power law could reflect this bounding. To test this hypothesis we compare the power law and the truncated power law:
 
 \begin{verbatim}
 > fit.distribution_compare('power_law', 'truncated_power_law')
 Assuming nested distributions
 (-0.081, 0.687)
 \end{verbatim}
 
 In fact, neither distribution is a significantly stronger fit ($p>.05$). From this we can conclude only moderate support for a power law, without ruling out the possibility of exponential truncation.
 
\begin{figure}[!ht]
\begin{center}
\includegraphics[width=4in]{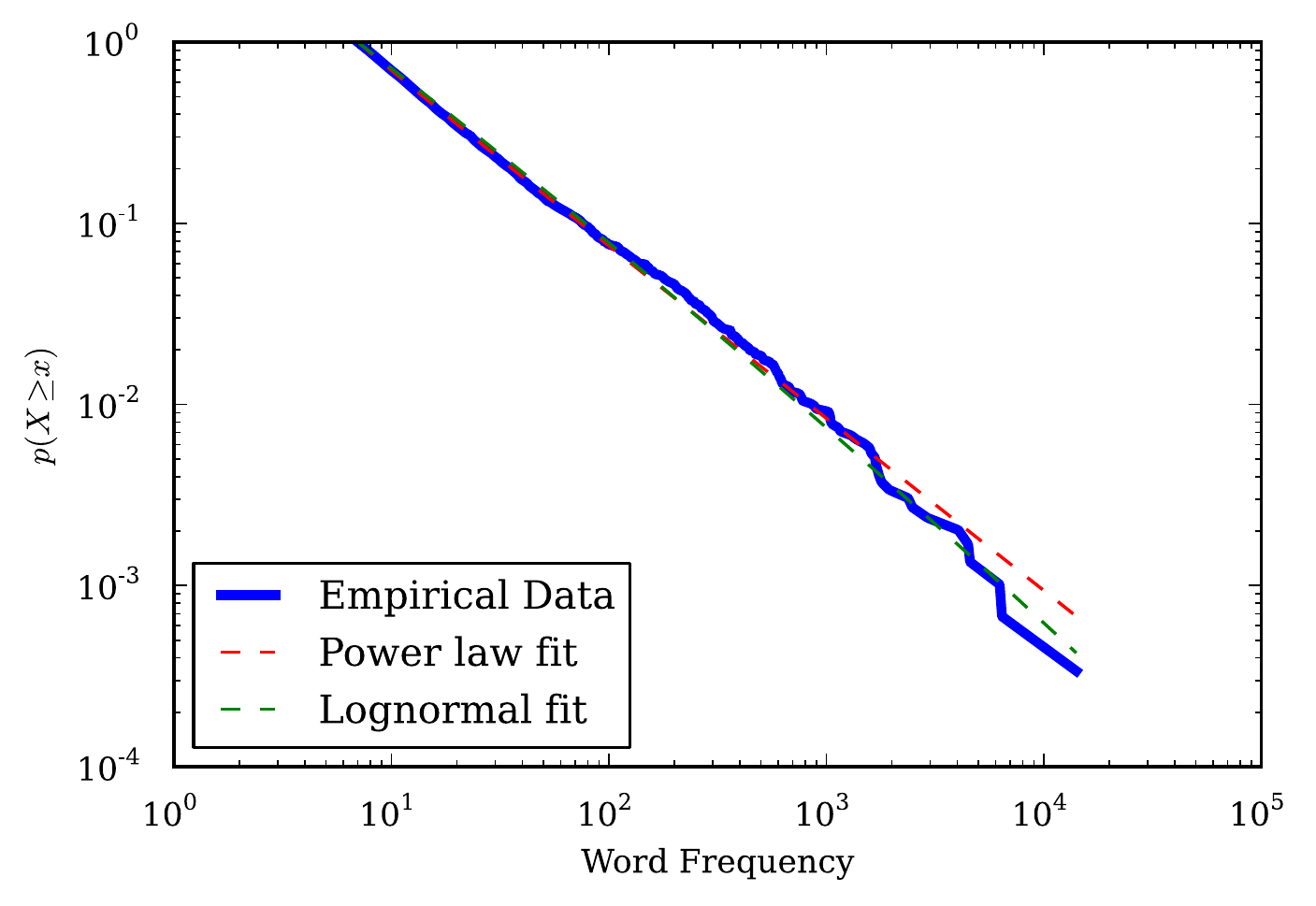}
\end{center}
\caption{
{\bf Complemenatary cumulative distribution functions of word frequency data and fitted power law and lognormal distributions.}
}
\label{Lognormal}
\end{figure}

 The importance of considering generative mechanisms is even greater when examining other heavy-tailed distributions. Perhaps the simplest generative mechanism is the accumulation of independent random variables, the central limit theorem. When random variables are summed, the result is the normal distribution. However, when positive random variables are multiplied, the result is the lognormal distribution, which is quite heavy-tailed. If the generative mechanism for the lognormal is plausible for the domain, the lognormal is frequently just as good a fit as the power law, if not better. Figure \ref{Lognormal} illustrates how the word frequency data is equally well fit by a lognormal distribution as by a power law ($p>.05$):
 
 \begin{verbatim}
 > fit.distribution_compare('power_law', 'lognormal')
 (0.928, 0.426)
 > fig4 = fit.plot_ccdf(linewidth=3)
 > fit.power_law.plot_ccdf(ax=fig4, color='r', linestyle='--')
 > fit.lognormal.plot_ccdf(ax=fig4, color='g', linestyle='--')
 \end{verbatim}
 
 There are domains in which the power law distribution is a superior fit to the lognormal (ex. \cite{Klaus2011}). However, difficulties in distinguishing the power law from the lognormal are common and well-described, and similar issues apply to the stretched exponential and other heavy-tailed distributions \cite{Malevergne2009, Malevergne2005,Mitzenmacher2004}. If faced with such difficulties it is important to remember the basic principles of hypothesis and experiment: Domain-specific generative mechanisms provide a basis for deciding which heavy-tailed distributions to consider as a hypothesis fit. Once candidates are identified, if the loglikelihood ratio test cannot distinguish between them the strongest solution is to construct an experiment to identify what generative mechanisms are at play. 
 
 \section*{Creating Simulated Data}
 Creating simulated data drawn from a theoretical distribution is frequently useful for a variety of tasks, such as modeling. Individual \verb$Distribution$ objects can generate random data points with the function \verb$generate_random$. These \verb$Distribution$ objects can be called from a \verb$Fit$ object or created manually.
 
 \begin{verbatim}
 > fit = powerlaw.Fit(empirical_data)
 > simulated_data = fit.power_law.generate_random(10000)
 
 > theoretical_distribution = powerlaw.Power_Law(xmin=5.0, parameters=[2.5])
 > simulated_data = theoretical_distribution.generate_random(10000)
 \end{verbatim}
 
 Such simulated data can then be fit again, to validate the accuracy of fitting software such as \verb$powerlaw$:
 
 \begin{verbatim}
 > fit = powerlaw.Fit(simulated_data)
 Calculating best minimal value for power law fit
 > fit.power_law.xmin, fit.power_law.alpha
 (5.30, 2.50)
 \end{verbatim}
 
 Validations of \verb$powerlaw$'s fitting of $\alpha$ and $x_{min}$ are shown on simulated power law data for a variety of parameter values in Figure S1.
 
 \section*{Advanced Considerations}
 
 \subsection*{Discrete Distribution Calculation and Estimation}
 While the maximum likelihood fit to a continous power law for a given $x_{min}$ can be calculated analytically, and thus the optimal $x_{min}$ and resulting fitted parameters can be computed quickly, this is not so for the discrete case. The maximum likelihood fit for a discrete power law is found by numerical optimization, the computation of which for every possible value of $x_{min}$ can take time. To circumvent this issue, \verb$powerlaw$ can use an analytic estimate of $\alpha$, from \cite{Clauset2009}, which can "give results accurate to about 1\% or better provided $x_{min} \ge 6$" when not using an $x_{max}$.  This \verb"estimate_discrete" option is True by default. Returning to the blackouts data:
 
 \begin{verbatim}
 > fit = powerlaw.Fit(data, discrete=True, estimate_discrete=True)
 Calculating best minimal value for power law fit
 > fit.power_law.alpha
   2.26912
 > fit.power_law.estimate_discrete
   True
 > fit = powerlaw.Fit(data, discrete=True, estimate_discrete=False)
 Calculating best minimal value for power law fit
 > fit.power_law.alpha
   2.26914
 > fit.power_law.estimate_discrete
   False
 \end{verbatim}
 
 Additionally, the discrete forms of some distributions are not analytically defined (ex. lognormal and stretched exponential). There are two available approximations of the discrete form. The first is discretization by brute force. The probabilities for all the discrete values between $x_{min}$ and a large upper limit are calculated with the continuous form of the distribution. Then the probabilities are normalized by their sum. The upper limit can be set to a specific value, or $x_{max}$, if present. The second approximation method is discretization by rounding, in which the continuous distribution is summed to the nearest integer. In this case, the probability mass at $x$ is equal to the sum of the continuous probability between $x-0.5$ through $x+0.5$. Because of its speed, this rounding method is the default.
 
 \begin{verbatim}
 > fit = powerlaw.Fit(data, discrete=True, xmin=230.0, xmax=9000, discrete_approximation='xmax')
 > fit.lognormal.mu
   -44.19
 > fit = powerlaw.Fit(data, discrete_approximation=100000, xmin=230.0, discrete=True)
 > fit.lognormal.mu
   0.28
 > fit = powerlaw.Fit(data, discrete_approximation='round', xmin=230.0, discrete=True)
 > fit.lognormal.mu
   0.40
 \end{verbatim}
 
 Generation of simulated data from a theoretical distribution has similar considerations for speed and accuracy. There is no rapid, exact calculation method for random data from discrete power law distributions. Generated data can be calculated with a fast approximation or with an exact search algorithm that can run several times slower \cite{Clauset2009}. The two options are again selected with the \verb$estimate_discrete$ keyword, when the data is created with \verb$generate_random$. 
 
 \begin{verbatim}
 > theoretical_distribution = powerlaw.Power_Law(xmin=5.0, parameters=[2.5], discrete=True)
 > simulated_data = theoretical_distribution.generate_random(10000, estimate_discrete=True)
 \end{verbatim}
 
 If the decision to use an estimation is not explictly assigned when calling \verb$generate_random$, the default behavior is to use the behavior used in the \verb$Distribution$ object generating the data, which may have been created by the user or created inside a \verb$Fit$ object.
 
 \begin{verbatim}
 > theoretical_distribution = powerlaw.Power_Law(xmin=5.0, parameters=[2.5], discrete=True, estimate_discrete=False)
 > simulated_data = theoretical_distribution.generate_random(10000)
 
 > fit = powerlaw.Fit(empirical_data, discrete=True, estimate_discrete=True)
 Calculating best minimal value for power law fit
 > simulated_data = fit.power_law.generate_random(10000)
 \end{verbatim}
 
 The fast estimation of random data has an error that scales with the $x_{min}$. When $x_{min}=1$ the error is over 8\%, but at $x_{min}=5$ the error is less than 1\% and at $x_{min}=10$ less than .2\% \cite{Clauset2009}. Thus, for distributions with small values of $x_{min}$ the exact calculation is likely preferred. 
 %
 %
 %
 %
 %
 %
 
 Random data generation methods for discrete versions of other, non-power law distributions all presently use the slower, exact search algorithm. Estimates of rapid, exact calculations for other distributions can later be implemented by users as they are developed, as described below.
 
 \subsection*{Nested Distributions}
 Comparing the likelihoods of distributions that are nested versions of each other requires a particular calculation for the resulting p-value \cite{Clauset2009}. Whether the distributions are nested versions of each other can be dictated with the \verb$nested$ keyword. If this keyword is not used, however, \verb$powerlaw$ automatically detects when one candidate distribution is a nested version of the other by using the names of the distributions as a guide. The appropriate corrections to the calculation of the p-value are then made. This is most relevant for comparing power laws to exponentially truncated power laws, but is also the case for exponentials to stretched exponentials (also known as Weibull distributions). 
 
 \begin{verbatim}
 > fit.distribution_compare('power_law', 'truncated_power_law')
 Assuming nested distributions
   (-0.3818, 0.3821)
 > fit.distribution_compare('exponential', 'stretched_exponential')
 Assuming nested distributions
   (-13.0240, 3.3303e-07)
 \end{verbatim}
 
 \subsection*{Restricted Parameter Range}\label{FittingMethodology}
 Each \verb$Distribution$ has default restrictions on the range of its parameters may take (ex. $\alpha > 1$ for power laws, and $\lambda > 0$ for exponentials). The user may also provide customized parameter ranges. A basic option is the keyword \verb$sigma_threshold$ (default \verb$None$), which restricts $x_{min}$ selection to those that yield a $\sigma$ below the threshold. 
 
 \begin{verbatim}
 > fit = powerlaw.Fit(data)
 Calculating best minimal value for power law fit
 > fit.power_law.alpha, fit.power_law.sigma, fit.xmin
   (2.27, 0.17, 230.00)
   
 > fit = powerlaw.Fit(data, sigma_threshold=.1)
 Calculating best minimal value for power law fit
 > fit.power_law.alpha, fit.power_law.sigma, fit.xmin
   (1.78, 0.06, 50.00)  
 \end{verbatim}
 
 More extensive parameter ranges can be set with the keyword \verb$parameter_range$, which accepts a dictionary of parameter names and a tuple of their lower and upper bounds. Instead of operating as selections on $x_{min}$ values, these parameter ranges restrict the fits considered for a given $x_{min}$.
 
 \begin{verbatim}
 > parameter_range = {'alpha': [2.3, None], 'sigma': [None, .2]}
 > fit = powerlaw.Fit(data, parameter_range=parameter_range)
 Calculating best minimal value for power law fit
 > fit.power_law.alpha, fit.power_law.sigma, fit.xmin
 (2.30, 0.17, 234.00)
 \end{verbatim}
 
 Even more complex parameter ranges can be defined by instead passing \verb$parameter_range$ a function, to do arbitrary calculations on the parameters. To incorporate the custom parameter range in the optimizing of $x_{min}$ the power law parameter range should be defined at initalization of the \verb$Fit$.
  
 \begin{verbatim}
 > parameter_range = lambda(self): self.sigma/self.alpha < .05
 > fit = powerlaw.Fit(data, parameter_range=parameter_range)
 Calculating best minimal value for power law fit
 > fit.power_law.alpha, fit.power_law.sigma, fit.xmin
   (1.88, 0.09, 124.00)  
 \end{verbatim}
 
 The other constituent \verb"Distribution" objects can be individually given a new parameter range afterward with the \verb$parameter_range$ function, as shown later.
 
 \subsection*{Multiple Possible Fits}
\begin{figure}[!ht]
\begin{center}
\includegraphics[width=4in]{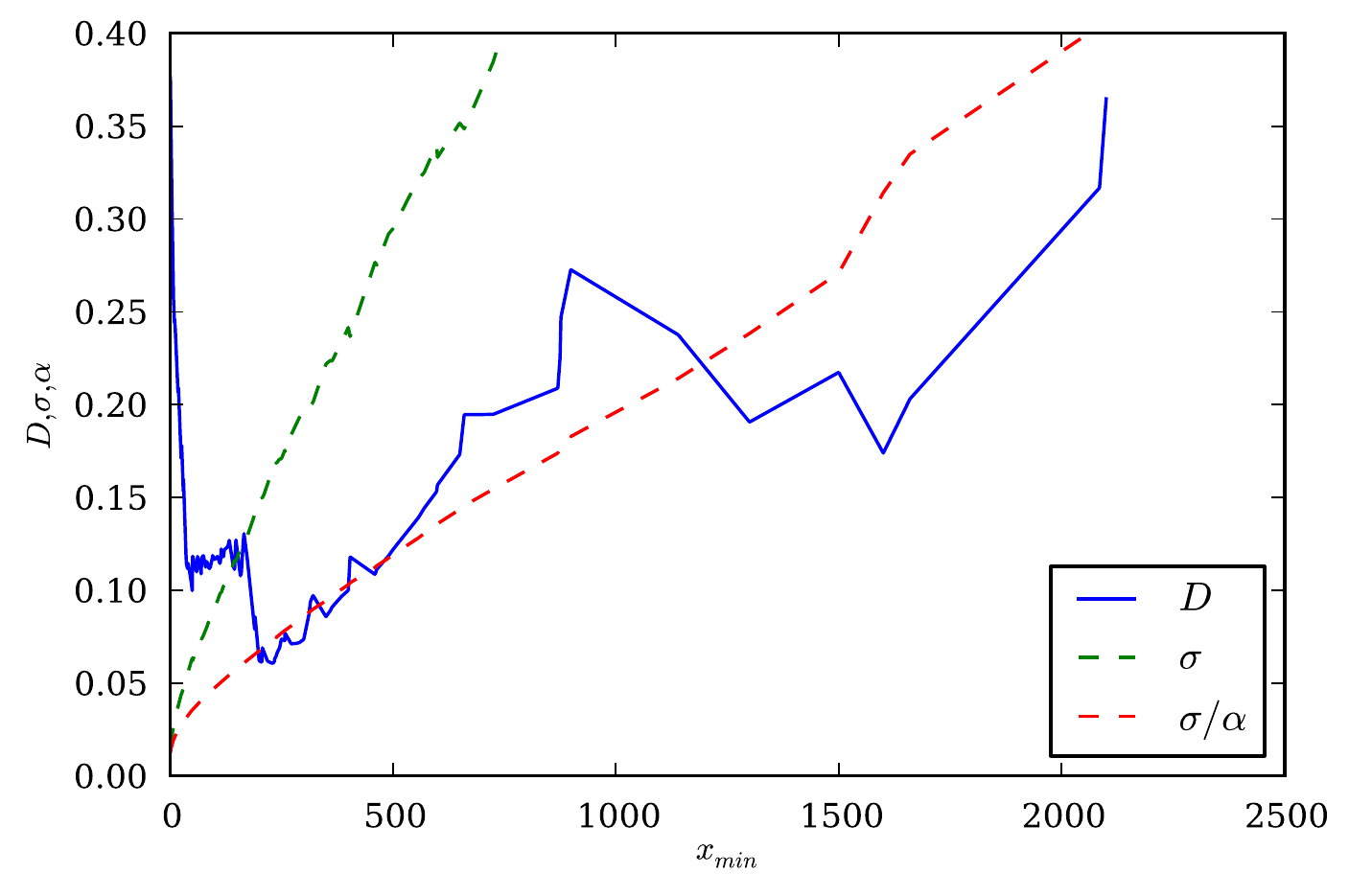}
\end{center}
\caption{
{\bf Example of multiple local minima of Kolmogorov-Smirnov distance $D$ across $x_{min}$.} As a power law is fitted to data starting from different $x_{min}$, the goodness of fit between the power law and the data is measured by the Kolmogorov-Smirnov distance $D$, with the best $x_{min}$ minimizing this distance. Here fitted data is the population sizes affected by blackouts. While there exists a clear absolute minima for $D$ at 230, and thus 230 is the optimal $x_{min}$ additional restrictions could exclude this fit. Parameter requirements such as $\sigma<.1$ or $\sigma /\alpha<.05$ would restrict the $x_{min}$ values considered, leading to the identification of a different, smaller $x_{min}$ at 50.
}
\label{Ds}
\end{figure}

 Changes in $x_{min}$ with different parameter requirements illustrate that there may be more than one fit to consider. Assuming there is no $x_{max}$, the optimal $x_{min}$ is selected by finding the $x_{min}$ value with the lowest Kolmogorov-Smirnov distance, $D$, between the data and the fit for that $x_{min}$ value. If $D$ has only one local minimum across all $x_{min}$ values, this is philosophically simple. If, however, there are multiple local minima for $D$ across $x_{min}$ with similar $D$ values, it may be worth noting and considering these alternative fits. For this purpose, the \verb$Fit$ object retains information on all the \verb$xmins$ considered, along with their \verb$Ds$, \verb$alphas$, and \verb$sigmas$. Returning to the data of population size affect by blackouts, Figure \ref{Ds} shows there are actually two values of $x_{min}$ with a local minima of $D$, and they yield different $\alpha$ values. The first is at $x_{min}=50$, and has a $D$ value of .1 and an $\alpha$ value of 1.78. The second, the more optimal fit, is $x_{min}=230$, with a $D$ of .06 and $\alpha$ of 2.27.
 
 \begin{verbatim}
 > from matplotlib.pylab import plot
 > plot(fit.xmins, fit.Ds)
 > plot(fit.xmins, fit.sigmas)
 > plot(fit.xmins, fit.sigmas/fit.alphas)
 \end{verbatim}
 
 The second minima may seem obviously optimal. However, $\sigma$ increases nearly monotonically throughout the range of $x_{min}$. If the user had included a parameter fitting requirement on $\sigma$, such as \verb$sigma_threshold=.1$, then the second, lower $D$ value fit from $x_{min}=230$ would not be considered. Even a more nuanced parameter requirement, such as $\sigma / \alpha < .05$, would exclude the second minimum. Which of the two fits from the two $x_{min}$ values is more appropriate may require domain-specific considerations. 
 
 \subsection*{No Possible Fits}
 When fitting a distribution to data, there may be no valid fits. This would most typically arise from user-specified requirements, like a maximum threshold on $\sigma$, set with \verb$sigma_threshold$. There may not be a single value for $x_{min}$ for which $\sigma$ is below the threshold. If this occurs, the threshold requirement will be ignored and the best $x_{min}$ selected. The \verb$Fit$ object's attribute \verb$noise_flag$ will be set to \verb$True$. 
 
 \begin{verbatim}
 > fit = powerlaw.Fit(data, sigma_threshold=.001)
 No valid fits found.
 > fit.power_law.alpha, fit.power_law.sigma, fit.xmin, fit.noise_flag
   (2.27, 0.17, 230.00, True)
 \end{verbatim}
 
 User-specified parameter limits can also create calculation difficulties with other distributions. Most other distributions are determined numerically through searching the parameter space from an initial guess. The initial guess is calculated from the data using information about the distribution's form. If an extreme parameter range very far from the optimal fit with a standard parameter range is required, the initial guess may be too far away and the numerical search will not be able to find the solution. In such a case the initial guess will be returned and the \verb$noise_flag$ attribute will also be set to \verb$True$. This difficulty can be overcome by also providing a set of initial parameters to search from, namely within the user-provided, extreme parameter range. 
 
 \begin{verbatim}
 > fit.lognormal.mu, fit.lognormal.sigma
   (0.15, 2.30)
 > range_dict = {'mu': [11.0, None]}
 > fit.lognormal.parameter_range(range_dict)
 No valid fits found.
 > fit.lognormal.mu, fit.lognormal.sigma, fit.lognormal.noise_flag
   (6.22, 0.72, True)
 
 > initial_parameters = (12, .7)
 > fit.lognormal.parameter_range(range_dict, initial_parameters)
 > fit.lognormal.mu, fit.lognormal.sigma, fit.lognormal.noise_flag
   (11.00, 5.72, False)
 \end{verbatim}

 \subsection*{Maximum Likelihood and Independence Assumptions}
 A fundamental assumption of the maximum likelihood method used for fitting, as well as the loglikelihood ratio test for comparing the goodness of fit of different distributions, is that individual data points are independent \cite{Clauset2009}. In some datasets, correlations between observations may be known or expected. For example, in a geographic dataset of the elevations of peaks, of the observation of a mountain of height $X$ could be correlated with the observation of foothills nearby of height $X/10$. Large correlations can potentially greatly alter the quality of the maximum likelihood fit. Theoretically, such correlations may  be incorporated into the likelihood calculations, but doing so would greatly increase the computational requirements for fitting.
 
 Depending on the nature of the correlation, some datasets can be "decorrelated" by selectively ommitting portions of the data \cite{Klaus2011}. Using the foothills example, the correlated foothills may be known to occurr within 10km of a mountain, and beyond 10km the correlations drops to 0. Requiring a minimum distance of 10km between observations of peaks, and ommitting any additional observations within that distance, would decorrelate the dataset. 
 
 An alternative to maximum likelihood estimation is minimum distance estimation, which fits the theoretical distribution to the data by minimizing the Kolmogorov-Smirnov distance between the data and the fit. This can be accomplished in the \verb$Fit$ object by using the keyword argument \verb$fit_method='KS'$ at initialization. However, the use of this option will not solve the problem of correlated data points for the loglikelihood ratio tests used in \verb$distribution_compare$.
 
\subsection*{Selecting $x_{min}$ with Other Distance Metrics}
The optimal $x_{min}$ is defined as the value that minimizes the Kolmogorov-Smirnov distance, $D$, between the empirical data and the fitted power law. This distance $D$, however, is notably insensitive to differences at the tails of the distributions, which is where most of a power law's interesting behavior occurs. It may be desirable to use other metrics, such as Kuiper or Anderson-Darling, which give additional weight to the tails when measuring the distance between distributions. In practice, the Kuiper distance $V$ does not perform notably better than the Kolmogorov-Smirnov distance \cite{Clauset2009}. The Anderson-Darling distance $A^{2}$ is actually so conservative that it can cut off too much of the data, leaving too few data points for a good fit \cite{Clauset2009}, though this may not be a concern for very large datasets that have a great many data points in the tail. If desired, \verb$powerlaw$ supports selecting $x_{min}$ with these other distances, as called by the \verb$xmin_distance$ keyword (default \verb$'D'$):

\begin{verbatim}
> fit = powerlaw.Fit(data, xmin_distance='D')
> fit = powerlaw.Fit(data, xmin_distance='V')
> fit = powerlaw.Fit(data, xmin_distance='Asquare')
\end{verbatim}

 \section*{The powerlaw Software}
 \subsection*{Availability and Installation}
 Source code and Windows installers of \verb$powerlaw$ are available from the Python Package Index, PyPI, at https://pypi.python.org/pypi/powerlaw. It can be readily installed with \verb$pip$:
 
 \begin{verbatim}
 pip install powerlaw
 \end{verbatim}
 
 Source code is also available on GitHub at https://github.com/jeffalstott/powerlaw and Google Code at https://code.google.com/p/powerlaw/.
 
 \subsection*{Dependencies}
The \verb$powerlaw$ Python package is implemented solely in Python, and requires the packages NumPy, SciPy, matplotlib, and mpmath. NumPy, SciPy and matplotlib are very popular and stable open source Python packages useful for a wide variety of scientific programming needs. SciPy development is supported by Enthought, Inc. and all three are included in the Enthought Python Distribution. Mpmath is required only for the calculation of gamma functions in fitting to the gamma distribution and the discrete form of the exponentially truncated power law. If the user does not attempt fits to the distributions that use gamma functions, mpmath will not be required. The gamma function calculations in SciPy are not numerically accurate for negative numbers. If and when SciPy's implementations of the gamma, gammainc, and gammaincc functions becomes accurate for negative numbers, dependence on mpmath may be removed.
 
 \section*{The Utility and Future of powerlaw}
 There have been other freely-available software for fitting heavy-tailed distributions \cite{Clauset2009, Ginsburg2012}. Here we describe differences between these packages' design and features and those of \verb$powerlaw$.
 
 
 As described in this paper, fitting heavy-tailed distributions involves several complex algorithms, and keeping track of numerous options and features of the fitted data set. \verb$powerlaw$ uses an integrated system of \verb$Fit$ and \verb$Distribution$ objects so that the user needs to interact with only a few lines of code to perform the full analysis pipeline. In other software this integration does not exist, and requires much more elaborate code writing by the user in order to analyze a dataset completely.
 
 In fitting data there are multiple families of distributions that the user may need or wish to consider: power law, exponential, lognormal, etc. And there are different flavors within each family: discrete vs. continuous, with or without an $x_{max}$, etc. \verb$powerlaw$ is currently unique in building in support for numerous distribution families and all the flavors within each one. And because of the integrated system, users do not need to do anything special or complicated to access any of the supported distributions. No other software package currently offers support for the same depth and breadth of probability distributions and subtypes as \verb$powerlaw$.
 
 Lastly, much existing software was not written for code maintenance or expansion. The code architecture of \verb$powerlaw$ was designed for easy navigation, maintenance and extensibility. As the source code is maintained in a git repository on GitHub, it is straightforward for users to submit issues, fork the code, and write patches. The most obvious extensions users may wish to write are additional candidate distributions for fitting to the data and comparing to a power law fit. All distributions are simple subclasses of the \verb$Distribution$ class, and so writing additional custom distributions requires only a few lines of code. Already users have submitted suggestions and written improvements to certain distributions, which were able to slot in seamlessly due to modularly-organized code. Such contributions will continue to be added to \verb$powerlaw$ in future versions.
 
 \section*{Acknowledgments}
 The authors would like to thank Andreas Klaus, Mika Rubinov and Shan Yu for helpful discussions. The authors also thank Andreas Klaus and the authors of \cite{Clauset2009} and \cite{Ginsburg2012} for sharing their code for power law fitting. Their implementations were a critical starting point for making \verb$powerlaw$.
 
This research was supported by the Intramural Research Program of the National Institute of Mental Health. The Behavioural and Clinical Neuroscience Institute, University of Cambridge, is supported by the Wellcome Trust and the Medical Research Council (UK). J.A. is supported by the NIH-Oxford-Cambridge Scholarship Program. E.B. is employed half-time by the University of Cambridge, UK, and half-time by GlaxoSmithKline (GSK).
-
 \bibliography{library}
 
 \setcounter{figure}{0}
 \renewcommand{\thefigure}{S\arabic{figure}}
 
 \begin{figure}[!ht]
 \begin{center}
\includegraphics[width=4in]{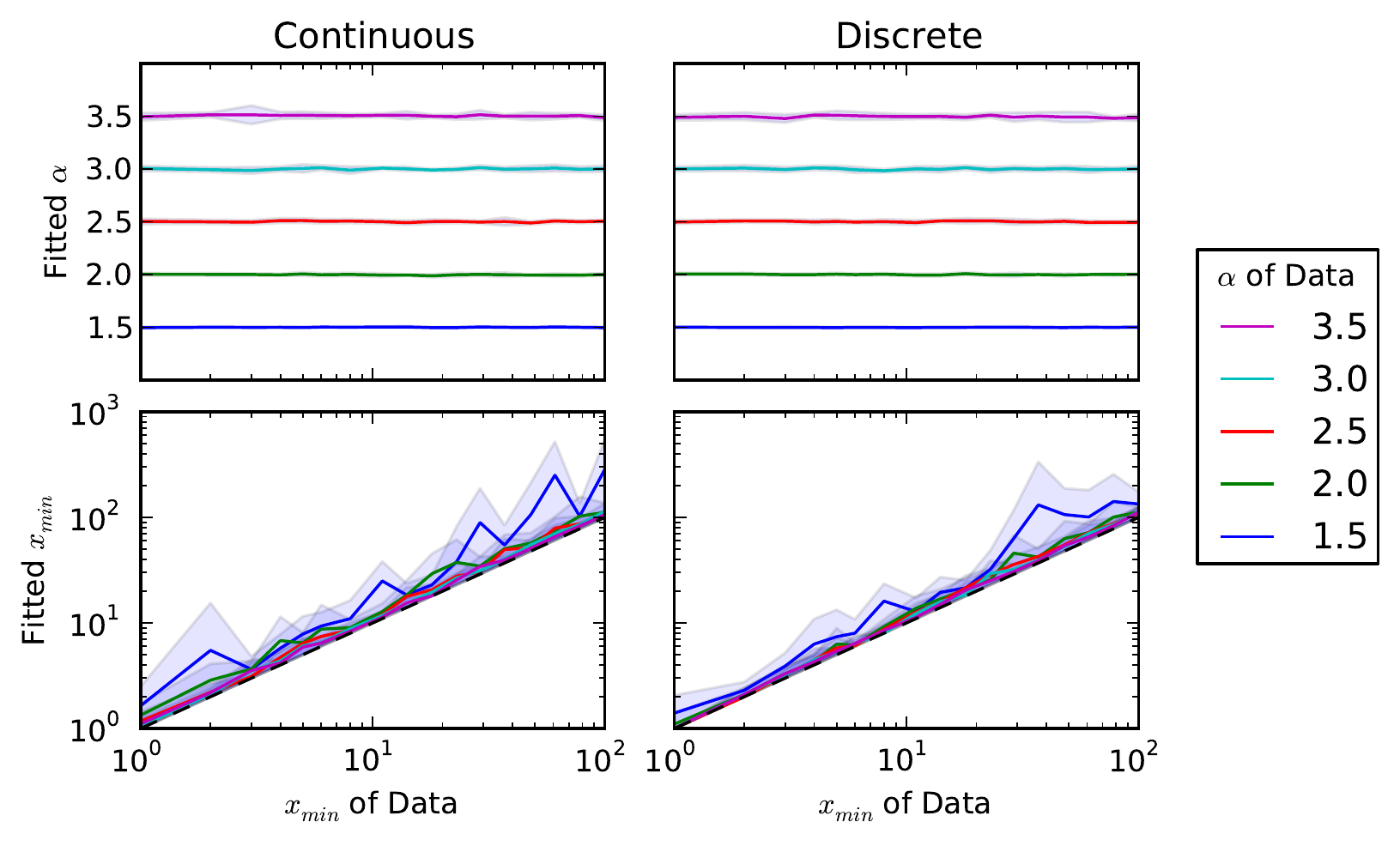}
 \end{center}
 \caption{
 {\bf Validation of fitting accuracy on simulated data with different values of $\alpha$ and $x_{min}$.} Each fit is the average of 10 simulated datasets of 10,000 data points each. Shading is the standard deviation of the 10 simulations. Note that on these simulated data there exist no data smaller than the true $x_{min}$ from which to sample, so any statistical fluctuation in the estimation of $x_{min}$ must return a value larger than the true value. The black dashed line on the bottom panels is the boundary where the fitted $x_{min}$ is equal to the actual $x_{min}$, below which fits cannot be made. For datasets in which there are noisy data below the $x_{min}$ of the power law, these methods recover the $x_{min}$ even more accurately, as shown in \cite{Clauset2009}.
 }
 \label{powerlaw_validation}
 \end{figure}
 
\setcounter{figure}{0}
\renewcommand{\figurename}{Code}

\begin{figure}
\begin{center}
\end{center}
\caption{
{\bf powerlaw code}. This version was used for all figures and examples. Future updates will be on the Python Package Index, Github and Google Code.
}
\end{figure}

\begin{figure}
\begin{center}
\end{center}
\caption{
{\bf Python code to make all figures.}
}
\end{figure}

\begin{figure}
\begin{center}
\end{center}
\caption{
{\bf Python code to make all figures, as IPython Notebook.}
}
\end{figure}

 \end{document}